\documentclass[10pt]{iopart}

\usepackage{epsfig,iopams}

\begin{document}

\title[Mortensen and Folkenberg, Low-loss criterion and effective area ....]{Low-loss criterion and effective area considerations for photonic crystal fibers}

\author{Niels Asger Mortensen and Jacob Riis Folkenberg} 

\address{Crystal Fibre A/S, Blokken 84, DK-3460 Birker\o d, Denmark}

\begin{abstract}
We study the class of endlessly single-mode all-silica photonic crystal fibers with a triangular air-hole cladding. We consider the sensibility to longitudinal nonuniformities and the consequences and limitations for realizing low-loss large-mode area photonic crystal fibers. We also discuss the dominating scattering mechanism and experimentally we confirm that both macro and micro-bending can be the limiting factor. 
\end{abstract}

\submitto{\JOA}

\section{Introduction}

\begin{figure}[t!]
\begin{center}
\epsfig{file=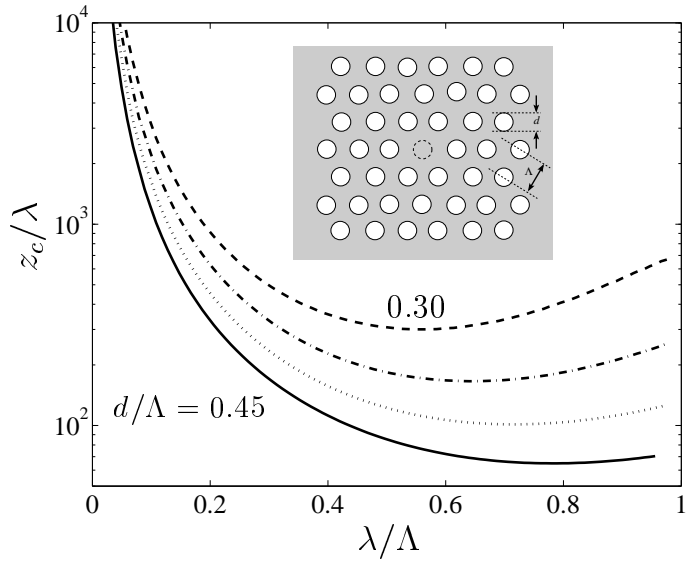, width=0.7\columnwidth,clip}
\end{center}
\caption{Plot of the coupling length $z_c/\lambda$ as a function of wavelength $\lambda/\Lambda$ for $d/ \Lambda = 0.3$, $0.35$, $0.4$, and $0.45$. The inset illustrates the dielectric structure in a cross section of the PCF.}
\label{fig1}
\end{figure}

Since the first results on photonic crystal fibers~\cite{knight1996,knight1997errata} (PCF) many exciting phenomena have been reported (for recent reviews see {\it e.g.} Refs.~\cite{bjarklev2001,birks2001,russell2003}). All-silica PCFs with a triangular air-hole cladding (see inset of Fig.~\ref{fig1}) have a big technological potential due to their endlessly single-mode properties~\cite{birks1997} combined with (in principle) unlimited large effective areas~\cite{knight1998_LMA}. However, it is well-known that scattering losses and confinement losses may put an important limit to the effective areas that can be realized~\cite{birks1997,sorensen2001,baggett2001,white2001}. In this paper we present very general results for the influence of longitudinal perturbations of the fiber axis on the loss and discuss how it correlates with the effective area $A_{\rm eff}$ and air-hole diameter. We also discuss important sources for the longitudinal perturbations and argue that macro-bending is not the only limiting factor in achieving low-loss fibers with large effective areas.

\section{Elastic scattering from the fundamental mode to the cladding}

The general phenomena of mode coupling from the fundamental mode to the cladding, which leads to attenuation of the light guided by the fiber, may be treated as an elastic scattering process. For single-mode fibers the scattering rate out of the mode can be expressed by a Fermi's golden rule expression (see {\it e.g.} Ref.~\cite{merzbacher})

\begin{equation}\label{rate}
\Gamma\propto\sum_m \big|W_m(\Delta\beta_m)\big|^2
\delta(\omega(\beta)-\omega(\beta_m)),
\end{equation}
where $\Delta\beta_m=\beta-\beta_m$ is the transferred momentum with $\{\beta_1,\beta_2,\ldots \}$ being cladding modes, {\it i.e.} modes not tightly confined to the core. Here, $W_m(\Delta\beta_m)$ is the matrix element between the unperturbed guided mode and the $m$th cladding mode for a given coupling mechanism. Scattering by longitudinal perturbations such as macro-bending, micro-bending, Bragg gratings, or dielectric imperfections are examples of elastic scattering processes covered by Eq.~(\ref{rate}). In Eq.~(\ref{rate}) a complete and orthogonal set of eigenmodes is assumed which is strictly speaking only fulfilled in the limit of vanishing absorption and confinement loss. However, also for the practical situation with weak absorption and confinement loss the analysis to follow is still quantitative correct as long as we for $\beta=\beta'+i\beta''$ have that $\beta''\ll \beta'$.

For the continuum of cladding modes there is a lower bound $\beta_{\rm cl}$ on their propagation constants, {\it i.e.} $\beta_m \leq \beta_{\rm cl}$. For a homogeneous cladding material of index $n_{\rm cl}$ one would typically have $\beta_{\rm cl}=n_{\rm cl}k$ and in a micro-structured cladding one can in a similar way define a fundamental space-filling mode \cite{broeng1999} with propagation constant $\beta_{\rm cl}$. In Eq.~(\ref{rate}) the finite $\Delta\beta\equiv \beta-\beta_{\rm cl}$ thus cuts off the effects of long-length-scale non-uniformities. From the corresponding coupling length (or beat length) 

\begin{equation}\label{zc}
z_c=2\pi/\Delta\beta
\end{equation}
and the characteristic length scale of a specific perturbation, $L_n$, a generalized low-loss criterion can thus be formulated~\cite{love} where losses will be significant only if 

\begin{equation}\label{high-loss}
\lambda/n_{\rm eff} \lesssim L_n \lesssim z_c.
\end{equation}
Otherwise, if $L_n < \lambda/n_{\rm eff}$, the rapid-varying perturbation is effectively reduced by the averaging over the wavelength, or if $L_n > z_c$ the perturbation is sufficiently slow to act as an adiabatic tapering-like process.
As emphasized in Ref.~\cite{love} the criterion in Eq.~(\ref{high-loss}) does not quantify loss, but it gives a correct parametric dependence of loss for various loss mechanisms. In order to quantify loss accurately one would need to know the matrix elements in Eq.~(\ref{rate}) for the full set of cladding modes and for all the scattering mechanisms at action. In this work we will rather use the criterion in Eq.~(\ref{high-loss}) to gain more general insight in the susceptibility of PCFs to longitudinal non-uniformities.

Recently, Eq.~(\ref{high-loss}) was used to identify improved large-mode area PCF designs~\cite{mortensen2003} and for analysis of micro-bending induced scattering loss~\cite{nielsen2003}. The formalism has also been used in a study of structural long-period gratings in PCFs with $L_n$ being the period of the gratings~\cite{kakarantzas2002}. Below we discuss three important
scattering mechanisms in more detail.

\subsection{Macro-bending losses}

For macro-bending loss it can be shown that~\cite{love}

\begin{equation}
\label{Ln_bending}
L_n^{\rm macro} \sim \big[2 R / \sqrt{\beta^2-\beta_{\rm cl}^2}\big]^{1/2}
\end{equation}
where $R$ is the bending radius. A critical bend-radius $R_c$ can then be introduced from $z_c\sim L_n$;

\begin{equation}\label{Rc}
R_c\sim \sqrt{\beta^2-\beta_{\rm cl}^2}/(\beta-\beta_{\rm cl})^2
\end{equation}
where we have omitted a numerical prefactor. While this expression may seem too simplistic we find that it fully accounts for our experimental data.

For PCFs the wavelength dependence of $\beta$ and $\beta_{\rm cl}$ (see {\it e.g.} Ref.~\cite{broeng1999}) gives rise to both a long-wavelength bend-edge (as in standard fibers) and also a short-wavelength bend-edge~\cite{birks1997,sorensen2001}. For the short-wavelength bend-edge Eqs.~(\ref{high-loss}) and (\ref{Ln_bending}) can be used to show that $R_c \propto \lambda^{-2}$~\cite{birks1997} which has been confirmed experimentally for a particular fiber design \cite{birks1997}.

\subsection{Micro-bending losses}

Micro-bending may be thought of as small and arbitrarily spaced bends of the fiber axis caused by external or frozen-in mechanical deformations. Typically, the power spectra of the deformations are very broadband, but because of the stiffness of the fiber there is an effective cut-off such that the fiber is only susceptible to deformations with periods, $L_n^{\rm micro}$, limited by~\cite{bjarklev1989}

\begin{equation}\label{Ln_micro}
L_n^{\rm micro} > r (\pi E_f/E_c)^{1/4}.
\end{equation}
Here, $r$ is the fiber radius and $E_f$ and $E_c$ are the Young's moduli of the fiber material and the coating material, respectively. The cut-off given above relies on a simplified model, but essentially states that it is hard to mechanically deform the fiber with a period shorter than the fiber diameter itself.

\subsection{Scattering by dielectric imperfections}

While macro and microbending induced deformations can be considered extrinsic sources, dielectric imperfections act as an intrinsic source which is present even when the fiber is not perturbed from the outside. Index variations of the silica, surface roughness at the air-silica interfaces, strain, and spatial variations in fiber diameter and air-hole diameters are examples of dielectric imperfections. From a fabrication point of view the goal is to limit index fluctuations {\it i)} to the sub-wavelength scale where {\it e.g.} Rayleigh scattering is a weak source of loss compared to the other mechanisms treated here and/or {\it ii)} to a long-scale adiabatic behavior, see Eq.~(\ref{high-loss}). As demonstrated recently it is possible to limit scattering by dielectric imperfections significantly by careful fabrication of the PCF~\cite{farr2002}.

\section{The effective area}

The spatial extension of the mode is of great importance to several applications. For non-linear applications one would like to have a very confined mode (transverse extension comparable to the wavelength) whereas for high-power applications one would typically prefer a large mode (transverse extension much larger than the wavelength) to avoid non-linear effects and material damage. In order to quantify the transverse extension we consider the effective area~\cite{agrawal_b,mortensen2002a}

\begin{equation}\label{Aeff}
A_{\rm eff}= \Big[\int d{\boldsymbol r}_\perp I({\boldsymbol r}_\perp)\Big]^2
\Big[\int d{\boldsymbol r}_\perp I^2({\boldsymbol r}_\perp)\Big]^{-1},
\end{equation}
where $I({\boldsymbol r}_\perp)$ is the transverse intensity distribution of the fundamental mode. For a Gaussian mode of width $w$ Eq.~(\ref{Aeff}) gives $A_{\rm eff}=\pi w^2$ and the intensity distribution in this type of PCF can be considered close to Gaussian~\cite{mortensen2002a,mortensen2002b}.

\begin{figure}[t!]
\begin{center}
\epsfig{file=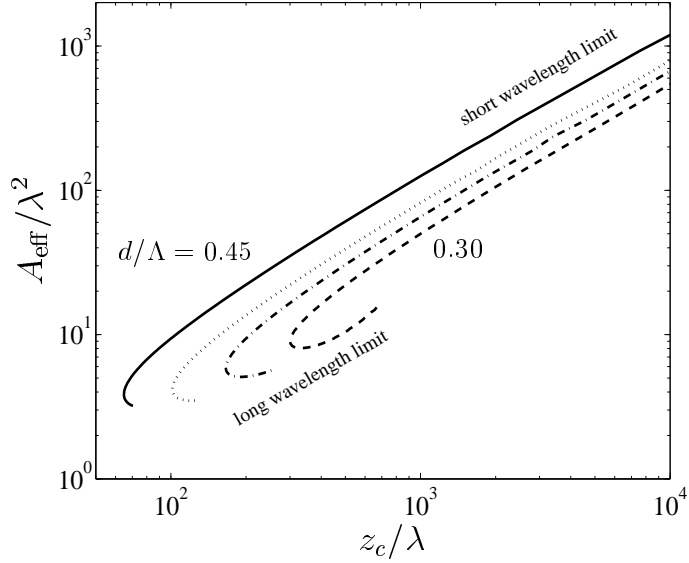, width=0.7\columnwidth,clip}
\end{center}
\caption{Plot of the effective area versus coupling length for $d/ \Lambda = 0.3$, $0.35$, $0.4$, and $0.45$.}
\label{fig2}
\end{figure}

\section{Numerical results}

The fully-vectorial eigenmodes of Maxwell's equations are computed with periodic boundary conditions in a planewave basis~\cite{johnson2000}. This approach provides the eigenmodes on a discrete lattice and the integrals in Eq.~(\ref{Aeff}) are then obtained by sums over lattice sites inside the super-cell. For the dielectric function we use $\varepsilon=1$ for the air holes and $\varepsilon=(1.444)^2 = 2.085$ for the silica. Ignoring the frequency dependence of the latter the wave equation becomes scale-invariant~\cite{joannopoulos} and all the results to be presented can thus be scaled to the desired value of $\Lambda$.

In Fig.~\ref{fig1} we show the coupling length, Eq.~(\ref{zc}), as a function of wavelength for different air-hole diameters. In general the coupling length decreases with increasing air-hole diameter indicating less scattering loss for increasing air-hole diameter. For $d/\Lambda=0.3$ the coupling length has a minimum at $\lambda/\Lambda \sim 0.5$. This minimum shifts toward higher wavelengths when increasing the hole diameter. For the endlessly single-mode limit $d/\Lambda\sim 0.45$ (see Ref.~\cite{kuhlmey} for a discussion of the particular value of this number) the minimum is at $\lambda/\Lambda \sim 0.8$. The occurrence of a minimum also indicates both a short wavelength and long wavelength loss edge as has been reported for macro-bending loss~\cite{birks1997,sorensen2001}. The position of the minimum confirms the results of Ref.~\cite{sorensen2001} where to a first approximation minimum bend-loss was found to occur at $\lambda\sim \Lambda/2$. The general picture is that when moving away from this minimum toward the large-mode area regime ($\lambda\ll \Lambda$) or the small-core (nonlinear) regime ($\lambda\sim \Lambda$) the PCF becomes more susceptible to longitudinal non-uniformities.
It is interesting to note that the recently reported PCF ($\Lambda\sim 4.2\,{\rm \mu m}$ and $d/\Lambda\sim 0.44$) by Farr {\it et al.}~\cite{farr2002} showed record-low loss at $\lambda=1550\,{\rm nm}$. Comparing to Fig.~\ref{fig1} the coupling length at the corresponding value $\lambda/\Lambda\sim 0.4$ is not very different from its minimum value (note the logarithmic scale) which indicates a close-to-minimal susceptibility to longitudinal non-uniformities for that fiber.

Calculating the effective area, Eq.~(\ref{Aeff}), as a function of wavelength~\cite{mortensen2002a} and combining the data with that of Fig.~\ref{fig1} we get the plot of effective area versus coupling length shown in Fig.~\ref{fig2}. For the short-wavelength limit ($\lambda \ll \Lambda$) we in general find that the coupling length increases with increasing effective area, {\it i.e.} for increasing effective area the PCF in general becomes more sensitive to scattering.

\begin{figure}[t!]
\begin{center}
\epsfig{file=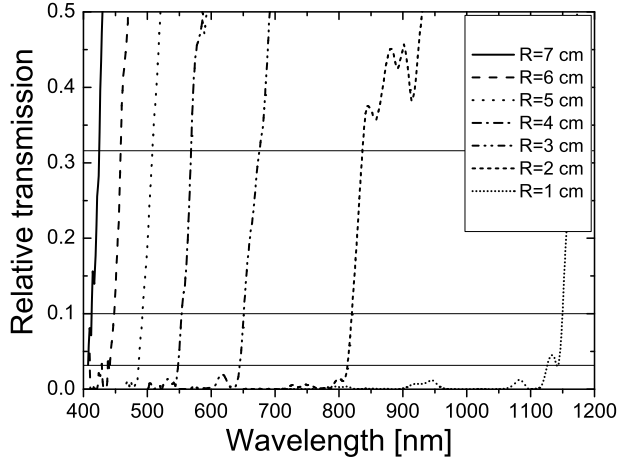, width=0.7\columnwidth,clip}
\end{center}
\caption{The transmission (linear scale) is plotted as a function of wavelength, for seven different bending radii of the fiber. The total bend of the fiber was 450 degrees on a dorn, and the transmission spectra were normalized to a straight fiber. The vertical lines correspond to a total decrease of the transmission of 5 dB, 10 dB and 15 dB respectively.}
\label{fig3}
\end{figure}

\begin{figure}[h!]
\begin{center}
\epsfig{file=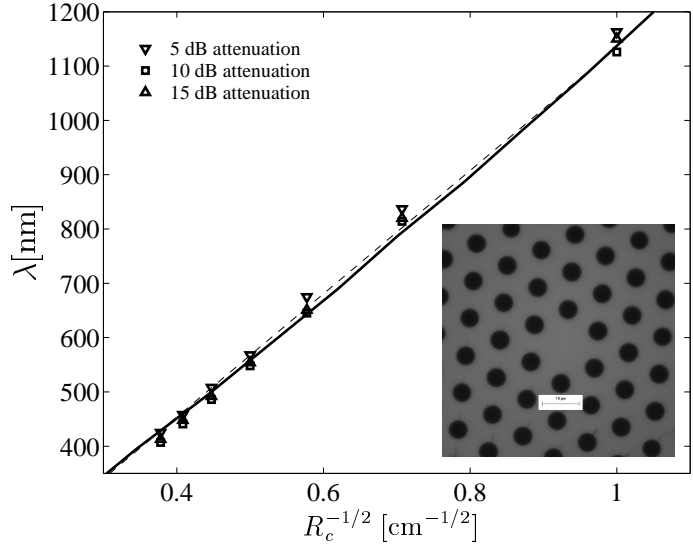, width=0.7\columnwidth,clip}
\end{center}
\caption{Critical bend radius versus wavelength. The data points are extracted from Fig.~\ref{fig3}, the dashed line shows a linear fit verifying the $R_c\propto \lambda^{-2}$ behavior, and the solid line shows $R_c(\lambda) \rightarrow \alpha\times R_c(\lambda)$ with $\alpha \simeq 1.8$ as fitting parameter, see Eq.~(\ref{Rc}).}
\label{fig4}
\end{figure}

\section{Experiments}
In order to experimentally quantify the different loss mechanisms described above we consider a PCF with $\Lambda \simeq 10\,{\rm \mu m}$ and $d/\Lambda\simeq 0.5$, see inset of Fig.~\ref{fig4}.

The macro bending losses of the fiber was characterized in a similar way to the first work on endlessly single-mode photonic crystal fibers~\cite{knight1996}. The fiber was coiled one and a quarter turn on seven dorns with different radii, and for each dorn the transmission spectrum was measured using a white light source and an optical spectrum analyzer. For all radii the launch of the light into the fiber was kept fixed, and the spectra were finally normalized with a reference spectrum obtained from the same piece of fiber kept straight. The spectra are shown in Fig.~\ref{fig3}.

The transmission spectra show a steep cut-off, which increases in wavelength as the bending radius decreases. For each radius we have measured the wavelengths corresponding to a 5 dB, 10 dB, and 15 dB transmission drop through the fiber, and as can be seen from Fig.~\ref{fig3} the three cut-off conditions occur within a very narrow wavelength range. In Fig.~\ref{fig4}, the cut-off wavelengths are plotted as a function of $R^{-1/2}$, and in agreement with previous observations~\cite{knight1996} a linear relationship is clearly observed, yielding a fit of (dashed line)

\begin{equation}
\lambda_c \simeq (1135 \pm 10\, nm)\times R_c^{-1/2}[{\rm cm}]
\end{equation}
for the particular fiber considered. The solid line shows Eq.~(\ref{Rc}) with the numerical prefactor adjusted to $\alpha \simeq 1.8$, {\it i.e.} $R_c(\lambda) \rightarrow \alpha\times R_c(\lambda)$. Thus, not only does our analysis predict the correct parametric behavior as in Ref.~\cite{knight1996} but a numerical prefactor of the order unity in Eq.~(\ref{Rc}) is also confirmed.

\begin{figure}[h!]
\begin{center}
\epsfig{file=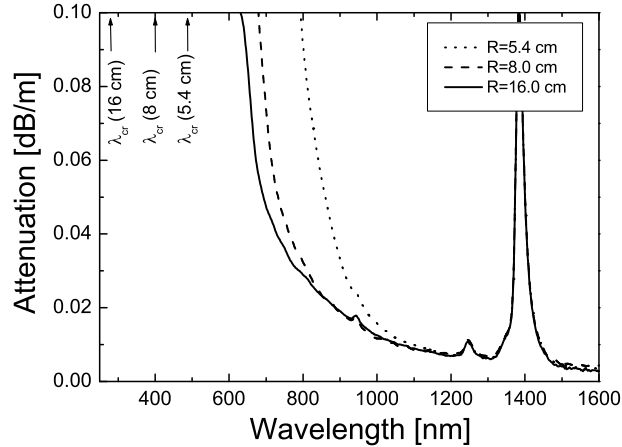, width=0.7\columnwidth,clip}
\end{center}
\caption{ The plot shows the spectral attenuation of the same fiber, wound on three different spool radii. The arrows indicate the critical wavelength for bending losses calculated using the fit shown in Fig.~\ref{fig4}.}
\label{fig5}
\end{figure}

Finally, in Fig.~\ref{fig5}, the spectral attenuation is shown for the same fiber wound on three different spool radii of 5.4 cm, 8 cm, and 16 cm. Since the fiber length is 100 m much smaller loss levels may be measured - typically with a precision of 0.001 dB/m. Also shown in Fig.~\ref{fig5} are the critical wavelengths for macro bending losses, calculated using the fitted relation above.

\subsection{Discussion}

The measurements show that in a wavelength range of some hundred nanometers above the critical wavelength for macrobending losses the attenuation in the fiber is significant. {\it E.g.} for a radius of 16 cm, the critical wavelength is 284 nm, but up to 620 nm the attenuation is still larger than 0.1 dB/m. This level is much larger than what would be expected theoretically from the tail of the macro bending edge \cite{sorensen2001}, but it is still dependent on the bending radius. Hence, the losses in this region are not due to material losses or scattering on point defects (as can be checked by OTDR measurements) but are attributed to micro-bending losses. Micro-bending studies as in Ref.~\cite{nielsen2003} confirm this. As the fiber is spooled on smaller diameters it is expected that the deformation spectrum experienced by the fiber will have contributions from shorter periods, which causes the microbending losses to extend to longer wavelengths where the beat length is shorter. Furthermore, the spectra coincide for wavelengths above $\approx$ 1200 nm, which may be related to the range where the fiber is no longer susceptible to external deformations. Indeed, the wavelength of 1200 nm corresponds to a beat length of $\approx$ 1 mm, which is in reasonable agreement with Eq.~(\ref{Ln_micro}). It should be pointed out that Eq.~(\ref{Ln_micro}) is based on a very
simple model of the mechanical deformation, which actually overestimates the susceptibility to external deformations because the so-called buffer effect of the coating is not taken into account~\cite{bjarklev1989}.

In fact the findings described above are very analogous to the properties found for standard fibers, only the wavelength dependence is reversed. Previously it was shown that the micro-bending sensitivity of standard fibers increases as the so-called MAC value increases~\cite{unger}. The MAC values is defined as ${\rm MFD}^2/\lambda_{\rm cut-off}$, which essentially is a numerical factor times the mode area divided by the index step between the core and the cladding. Since the index step is constant for a standard solid glass fiber and the mode area increases for increasing wavelengths, the MAC value also increases for increasing wavelength. However, for the type of PCFs described here, the mode area is practically constant but the effective index step decreases for decreasing wavelength, which implies an increasing MAC value for decreasing wavelength.

\section{Conclusion}

In this paper we have studied, both theoretically and experimentally, the loss mechanisms due to longitudinal perturbations of all-silica photonic crystal fibers with a triangular cladding structure. In the endlessly single-mode limit, the coupling length $z_c$ between the fundamental mode and the first cladding mode has been calculated for a range of hole sizes, and $z_c$ has been related to both the effective area, the critical radius for macrobending, and the sensitivity to microbending. It is found that a minimum coupling length exists, where the fiber is least susceptible to longitudinal perturbations. The position of the minimum is dependent on the hole size and varies from $\lambda/\Lambda \approx 0.5$ for $d/\Lambda = 0.3$ to $\lambda/\Lambda \approx 0.8$ for $d/\Lambda = 0.45$, while the minimum value of $z_c$ decreases by almost an order of magnitude from $d/\Lambda = 0.3$ to $d/\Lambda = 0.45$. In general $z_c$ increases for increasing $A_{\rm eff} / \lambda^2$, giving rise to a macrobending edge for short wavelengths as well as a wavelength range where the fiber is potentially susceptible to microbends on a length scale of several fiber diameters, which may easily be experienced in a practical fiber device or cable. Both the macro- and microbending losses are demonstrated experimentally, and a very good agreement with the calculated coupling lengths are found.

The results reported here allow for a comparison between single-mode fibers in the conventional all-glass technology and the endlessly single-mode photonic crystal fibers, where especially two points should be stressed. First of all, the low loss criterion formulated in Ref.~\cite{love} based on the coupling length between the fundamental mode and the cladding also holds for photonic crystal fibers. In practice however, the wavelength scale is reversed compared to standard fibers, so both macro {\em and} microbending losses are experienced at short wavelengths. Secondly, it has previously been concluded that the limiting factor for achieving large mode areas in endlessly single-moded PCFs is the macrobending losses~\cite{knight1998_LMA}, however the results shown here demonstrate that the sensitivity to microbending may be a serious limiting factor too. The microbending losses may be minimized {\it e.g.} by choosing the largest hole sizes for single-mode operation, a larger outer diameter of the fiber, or by improving the fiber production process to avoid frozen-in perturbations, but at present the lower limit is not well-known. Eventually, when microbending losses have been reduced to a minimum, it seems feasible to guide single-mode light at 400 nm with a MFD of 10 $\rm \mu m$ and bending radii down to 80 mm. A similar step-index fiber with a cut-off at 400 nm, would require a well-controlled index step of the order $\Delta n = 3\times 10^{-4}$, which requires a very accurate control of the dopant levels in the preform and limited dopant-diffusion during fiber pulling. While this is very difficult the same effective index step can be realized in PCFs with relative ease through control of air hole-sizes.

\section*{Acknowledgments}

We acknowledge useful discussions with M.~D. Nielsen (Crystal Fibre A/S and Research Center COM, Technical University of Denmark).

\end{document}